\documentclass{iopart}
\usepackage{epsfig}
\usepackage{amssymb}

\begin{document} 

\newcommand{\vk}{{\vec k}} 
\newcommand{\vK}{{\vec K}}  
\newcommand{\vb}{{\vec b}}  
\newcommand{\vp}{{\vec p}}  
\newcommand{\vq}{{\vec q}}  
\newcommand{\vQ}{{\vec Q}} 
\newcommand{\vx}{{\vec x}} 
\newcommand{\vh}{{\hat{v}}} 
\newcommand{\cO}{{\cal O}}
\newcommand{\be}{\begin{equation}} 
\newcommand{\ee}{\end{equation}}  
\newcommand{\half}{{\textstyle\frac{1}{2}}}  
\newcommand{\gton}{\mathrel{\lower.9ex \hbox{$\stackrel{\displaystyle 
>}{\sim}$}}}  
\newcommand{\lton}{\mathrel{\lower.9ex \hbox{$\stackrel{\displaystyle 
<}{\sim}$}}}  
\newcommand{\ben}{\begin{enumerate}}  
\newcommand{\een}{\end{enumerate}} 
\newcommand{\bit}{\begin{itemize}}  
\newcommand{\eit}{\end{itemize}} 
\newcommand{\bc}{\begin{center}}  
\newcommand{\ec}{\end{center}} 
\newcommand{\bea}{\begin{eqnarray}}  
\newcommand{\eea}{\end{eqnarray}}

\title{Meson and baryon elliptic flow at high $p_T$ from parton coalescence}
 
\date{\today}
 
\author{D\'enes Moln\'ar}
\address{Department of Physics, Ohio State University, Columbus, OH 43210}

%\ead{molnard@mps.ohio-state.edu}

\begin{abstract}
The large and saturating differential elliptic flow 
$v_2(p_\perp)$ observed in $Au+Au$ reactions at RHIC
so far could only be explained assuming 
an order of magnitude denser initial parton system than estimated from 
perturbative QCD.
Hadronization via parton coalescence can resolve this ``opacity puzzle''
because it enhances hadron elliptic flow at 
large $p_\perp$ relative to that of partons
at the same transverse momentum. 
An experimentally testable consequence of the coalescence scenario is that
$v_2(p_\perp)$ saturates
at about 50\% higher values for baryons than for mesons.
In addition, if strange quarks have weaker flow than light quarks,
hadron $v_2$ at high $p_\perp$ decreases with relative strangeness
content.
\end{abstract}


%\maketitle 

\section{Introduction and summary}
Differential elliptic flow, 
$v_2(p_\perp)\equiv\langle \cos(2\phi)\rangle_{p_\perp}$,
 the second Fourier moment of the azimuthal momentum distribution
for a given $p_\perp$, 
is one of the important experimental probes of collective
dynamics in $A+A$ reactions\cite{flow-review}.
Measurements of elliptic flow at high transverse momentum
provide important constraints
about the opacity of the partonic medium in heavy-ion collisions\cite{pQCDv2,v2,coalv2}.

\begin{figure}[hbpt] 
\begin{center}
\hspace*{-0.2cm}\epsfig{file=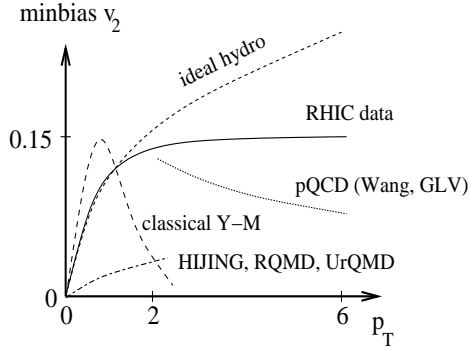,height=1.8in,width=2.4in,clip=5,angle=0} 
\end{center}
\vspace*{-0.2cm} 
\caption{\label{fig1}
Illustrative comparison between $v_2(p_\perp)$ data from RHIC and model calculations\cite{pQCDv2,CGCv2,Heinzhydro,Kolbhydro,Teaneyhydro,Kolbhighptv2,stringv2}.
}
\end{figure}

Recent data from RHIC for Au+Au reactions at $\sqrt{s_{NN}}=130$ and $200$~GeV
show a remarkable saturation property of elliptic flow 
for $\sim 2$ GeV $< p_\perp < \sim 6$ GeV with $v_2$ 
reaching up to 0.2 \cite{PHENIXv2,STARv2,KirillQM2002,PHENIXidentv2},
a value that corresponds to a factor of two azimuthal angle asymmetry
of  particle production relative to the reaction plane.
As illustrated in Fig.~\ref{fig1},
the saturation pattern
differs qualitatively from calculations based on a variety of
different models:
ideal (nondissipative) 
hydrodynamics\cite{Heinzhydro,Kolbhydro,Teaneyhydro,Kolbhighptv2},
hadronic cascades with string dynamics\cite{stringv2},
inelastic parton energy loss\cite{pQCDv2},
and classical Yang-Mills evolution\cite{CGCv2}.

\vskip 0.1cm
On the other hand, 
covariant parton transport theory\cite{ZPC,ZPCv2,nonequil,inelv2,v2,hbt}
is able to reproduce the observed flow saturation pattern.
The main ingredient of this theory is a finite local mean free path 
$\lambda(s,x) \equiv 1/\sigma(s) n(x)$,
providing a natural interpolation between
free streaming ($\lambda=\infty$)
and  ideal hydrodynamics ($\lambda=0$).
Several studies show
that initial parton densities
and elastic $2\to 2$ parton cross sections estimated from perturbative QCD,
$dN_g/d\eta(b=0) \sim 1000$ \cite{EKRT} and $\sigma_{gg\to gg}\approx 3$ mb,
generate too small collective effects at RHIC\cite{ZPCv2,inelv2,v2,hbt}.
Nevertheless, quantitative agreement with the $v_2(p_\perp)$ data 
is possible, provided initial parton densities and/or cross sections
are enhanced by an order of magnitude
to $\sigma dN_g/d\eta(b=0) \sim 45\,000$ mb\cite{v2}.
An enhancement of parton opacities seems necessary to explain 
the pion interferometry data from RHIC as well\cite{hbt}.

To compare to the experiments,
parton transport models have to incorporate 
the hadronization process.
The studies mentioned above considered two simple schemes:
1) $1 parton\to 1\pi$ hadronization, motivated by parton-hadron duality,
and 2) independent fragmentation.
An alternative is parton coalescence\cite{ALCORMICOR},
which became the focus of recent theoretical 
interest\cite{Voloshincoal,LinKov2,coalMtoB,coalv2,charmv2} 
as a possible explanation for the anomalous meson/baryon ratio and 
features of the elliptic flow data  at RHIC.
These studies indicate that hadron production at RHIC may be dominated by
coalescence out to surprisingly large transverse momenta
$p_\perp \sim 5$ GeV.

In this talk I 
show that hadronization via parton coalescence
can resolve the opacity puzzle
because it leads to an amplification of elliptic flow\cite{coalv2} at high 
$p_\perp$ (see Fig.~\ref{fig2}a).
With this enhancement,
parton transport calculations\cite{v2} 
can account for the charged particle elliptic flow data from RHIC
with
moderate initial parton densities and cross sections 
$dN_g/d\eta \sim 1500-3000$, $\sigma_{gg\to gg} = 3$ mb.

Parton coalescence predicts a rather simple relationship, Eq.~(\ref{v2LO}),
between parton flow coefficients
and those of hadrons.
If all partons have similar elliptic flow,
elliptic flow is stronger for baryons than for mesons,
$v_2^B \approx  1.5 v_2^M$, at high $p_\perp > 2-3$ GeV,
as shown in Fig.~\ref{fig2}a.
If on the other hand strange quarks show weaker flow than light quarks,
elliptic flow at high $p_\perp$ is ordered 
by relative strangeness content, such that
$v_2^p > v_2^\Lambda \approx v_2^\Sigma > v_2^\pi > v_2^K > v_2^\phi$,
$v_2^{\Lambda,\Sigma} > v_2^\Xi > v_2^K$, and $v_2^\Xi > v_2^\Omega 
\approx 3 v_2^\phi/2$ (see Fig.~\ref{fig2}b). 
These predictions are reinforced by preliminary
calculations using the MPC parton transport model\cite{MPC} 
(see Fig.~\ref{fig3}a) and 
can be readily tested in current and future heavy-ion collision experiments.

With the help of (\ref{v2LO}), one can also determine the elliptic flow
of the various parton species from $v_2$ measurements for several hadrons,
providing valuable insight on the evolution of the dense parton medium 
in heavy-ion collisions.

\section{Parton coalescence}

A convenient framework to start with is the local variant of the
coalescence model\cite{Voloshincoal,coalv2},
in which the coalescing partons have equal momenta.
This is justified if hadron wave functions are narrow in momentum
space and if constituent partons have similar masses.
See Ref.~\cite{charmv2} for a study of the more 
general case.

Assuming that all partons have the same momentum distribution, 
the hadron spectra at midrapidity are then given by
\be
\fl
\frac{dN_B}{d^2 p_\perp}(\vp_\perp) 
 = C_B(p_\perp) \left[  \frac{dN_q}{d^2 p_\perp}(\vp_\perp/3) \right]^3, \qquad
\frac{dN_M}{d^2 p_\perp}(\vp_\perp) 
 = C_M(p_\perp) \left[  \frac{dN_q}{d^2 p_\perp}(\vp_\perp/2) \right]^2 \ , 
\qquad
\label{coal_eq}
\ee
where $C_M$ and $C_B$ 
are the probabilities for
$q\bar{q}\to meson$ and $qqq \to baryon$ coalescence.
The coefficients $C$ can in general depend on momentum, for example,
if there is a strong radial flow.

Equation (\ref{coal_eq}) is valid for rare processes only. 
At high constituent phase space densities,
most quarks recombine into hadrons and 
the number of hadrons is {\em linearly} proportional to 
that of quarks, $dN_h(p_\perp) \propto dN_q(p_\perp/n)$ ($n=2,3$ for mesons, baryons).

At lower constituent densities, parton coalescence is
relatively rare and Eq.~(\ref{coal_eq}) is applicable.
Clearly, most partons hadronize via independent fragmentation in this case.
Still, hadron production can be dominated by coalescence in
a broad region of {\em hadron} transverse momenta.
The reason is that in independent fragmentation
hadrons with a given $p_\perp$ come from rare high-momentum partons
with momentum $p_\perp/z$ $(z<1)$, i.e.,
$dN_h^{frag}(p_\perp) \sim dN_q(p_\perp/z)$,
whereas via coalescence they can come from more abundant low-momentum partons 
with momenta $p_\perp/n$, i.e.,
$dN_h^{coal}(p_\perp) = C_h [dN_q(p_\perp/n)]^n$.

At very high transverse momenta,
the fragmentation process wins 
because due to the quadratic/cubic dependence in Eq.~(\ref{coal_eq})
the coalescence yield drops steeper as a function of $p_\perp$ than
the fragmentation yield.
E.g., a power law parton 
spectrum $dN_q/p_\perp dp_\perp \sim A p_\perp^{-\alpha}$
implies $dN_h^{coal}/dN_h^{frag} \sim C_h A^{n-1} p_\perp^{-(n-1)\alpha}
\to 0$ at high $p_\perp$.
The boundary between the coalescence and fragmentation regions may be 
at rather large transverse momenta $p_\perp \sim 5$ GeV\cite{coalMtoB}.

\section{Elliptic flow amplification and ordering}

In the coalescence region, meson and baryon elliptic flow are given
by that of partons via\cite{coalv2}
\be
v_{2,M}(p_\perp) \approx  2 v_{2,q} (\frac{p_\perp}{2}) , \quad
v_{2,B}(p_\perp) \approx  3 v_{2,q} (\frac{p_\perp}{3}) \ ,
\label{v2LO}
\ee
as follows from Eq.~(\ref{coal_eq}) and $v_2 \ll 1$.
For example, if partons have only elliptic anisotropy,
i.e., $dN_q/p_\perp dp_\perp d\Phi = (1/2\pi) dN_q/p_\perp dp_\perp [1+2v_{2,q} \cos (2\Phi)]$, then
\be
\fl
 v_{2,B}(p_\perp)  
  = \frac{3 v_{2,q} (p_\perp/3)+3v_{2,q}^3(p_\perp/3)}{1+6 v_{2,q}^2 (p_\perp/3)}, \qquad
 v_{2,M}(p_\perp)  = \frac{2 v_{2,q} (p_\perp/2)}{1+2 v_{2,q}^2 (p_\perp/2)} 
\ .
\ee
Note, Eq.~(\ref{v2LO}) is also valid for all flow coefficients $v_k\equiv \langle \cos k\phi\rangle$, 
not only $v_2$.

\begin{figure}[hbpt] 
\center
\hspace*{-0.2cm}\epsfig{file=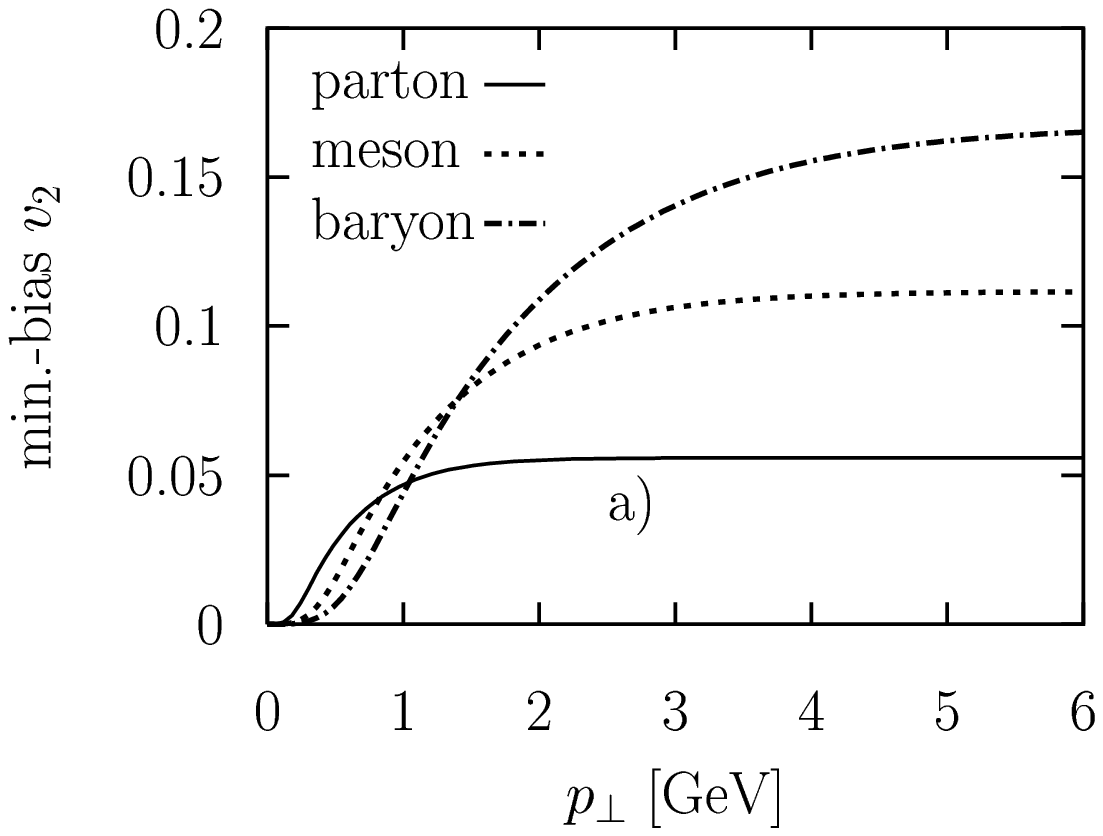,height=1.8in,width=2.2in,clip=5,angle=0} 
\epsfig{file=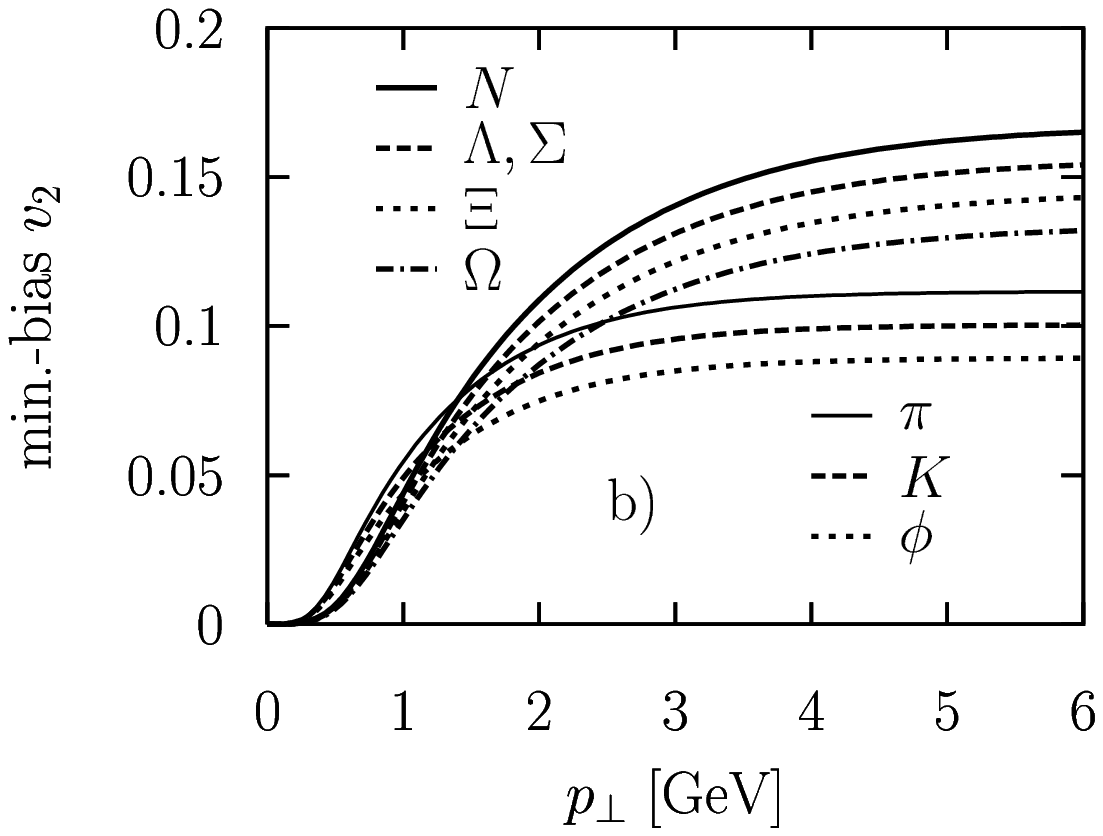,height=1.8in,width=2.2in,clip=5,angle=0}
\vspace*{-0.2cm} 
\caption{\label{fig2} 
Qualitative behavior of baryon and meson elliptic flow as a function of $p_\perp$ from parton coalescence, a) if all partons have the same $v_2$, b) if $v_2^s < v_2^q$.
}
\end{figure}

Fig.~\ref{fig2}a illustrates the effect of parton coalescence on 
baryon and meson elliptic flow, as given by Eq.~(\ref{v2LO}),
compared to parton elliptic flow.
The latter is shown schematically by the solid line.
At small transverse momenta, parton $v_2(p_\perp)\propto p_\perp^2$, 
as follows from general analyticity considerations. 
This region, before $v_2$ becomes approximately linear in $p_\perp$ 
could be relatively small
(depending on the effective mass of partons).
At higher transverse momenta $p_\perp > 1-2$~GeV,   
parton elliptic flow saturates as predicted by the MPC parton transport 
model\cite{v2}.  

In the low-$p_\perp$ region where $v_2(p_\perp)$ increases
faster than linearly, $v_{2,B}<v_{2,M}< v_{2,q}$. 
Though here Eq.~(\ref{v2LO}) may not be applicable,
it is nevertheless interesting that 
the data do exhibit such a behavior.
This ordering follows naturally from hydrodynamics,
where flow decreases with increasing
particle mass\cite{Heinzhydro,Kolbhydro,Teaneyhydro}.
Similar mass dependence could also arise in a coalescence model
because heavier hadrons can be formed by quarks with
larger relative momentum (ignored in the local coalescence approach).

On the other hand, at high $p_\perp$, where parton
$v_2(p_\perp)$ increases slower than linearly, 
baryon flow becomes larger than meson flow,
$v_{2,B}>v_{2,M} > v_{2,q}$, by as much as 50\%.    
Parton collective flow saturation above $p_\perp > 1-2$~GeV
results in saturating meson/baryon flow at  $p_\perp > 2-4$~GeV  that is
{\em amplified two/three-fold}  compared to that of partons.
Saturation sets in at $50\%$ higher $p_\perp$ for 
baryons than for mesons.
Note that any eventual decrease of parton elliptic flow
at very high $p_\perp$ (e.g., because of parton energy loss)
would happen at two to three times larger $p_\perp$ for hadrons.

The high-$p_\perp$ results above strongly differ from those 
obtained in Ref.~\cite{LinKov2}.
The reason is that, unlike Eq.~(\ref{v2LO}), in Ref.~\cite{LinKov2} the
coalescence of quarks was considered to be independent of their relative
momenta and therefore hadron elliptic flow at high
$p_\perp$ was similar to that of a high-$p_\perp$ quark.   

If not all partons show the same elliptic flow, further
 differentiation occurs because in that case
\bea
v_{2,B=abc}(p_\perp) &\approx& v_{2,a}(p_\perp/3) + v_{2,b}(p_\perp/3) +v_{2,c}(p_\perp/3)
\nonumber\\
v_{2,M=\bar{a}b}(p_\perp) &\approx& v_{2,\bar{a}}(p_\perp/2) + v_{2,b}(p_\perp/2)\ .
\label{v2LOdiff}
\eea
For example, strange quarks may have a smaller $v_2(p_\perp)$ 
than light quarks, at high $p_\perp$ because heavy quarks are expected
 to lose less energy 
in nuclear medium\cite{heavyqEL},
while at low $p_\perp$ due to the mass dependence of hydrodynamic flow.
As illustrated in Fig.~\ref{fig2}b, if $v_2^s < v_2^q$, 
elliptic flow decreases with increasing relative 
strangeness content within the baryon and meson bands,
i.e., $v_2^p > v_2^\Lambda \approx v_2^\Sigma > v_2^\Xi > v_2^\Omega$ 
and $v_2^\pi > v_2^K > v_2^\phi$. 

\begin{figure}[hbpt] 
\center
\hspace*{-0.2cm}
\epsfig{file=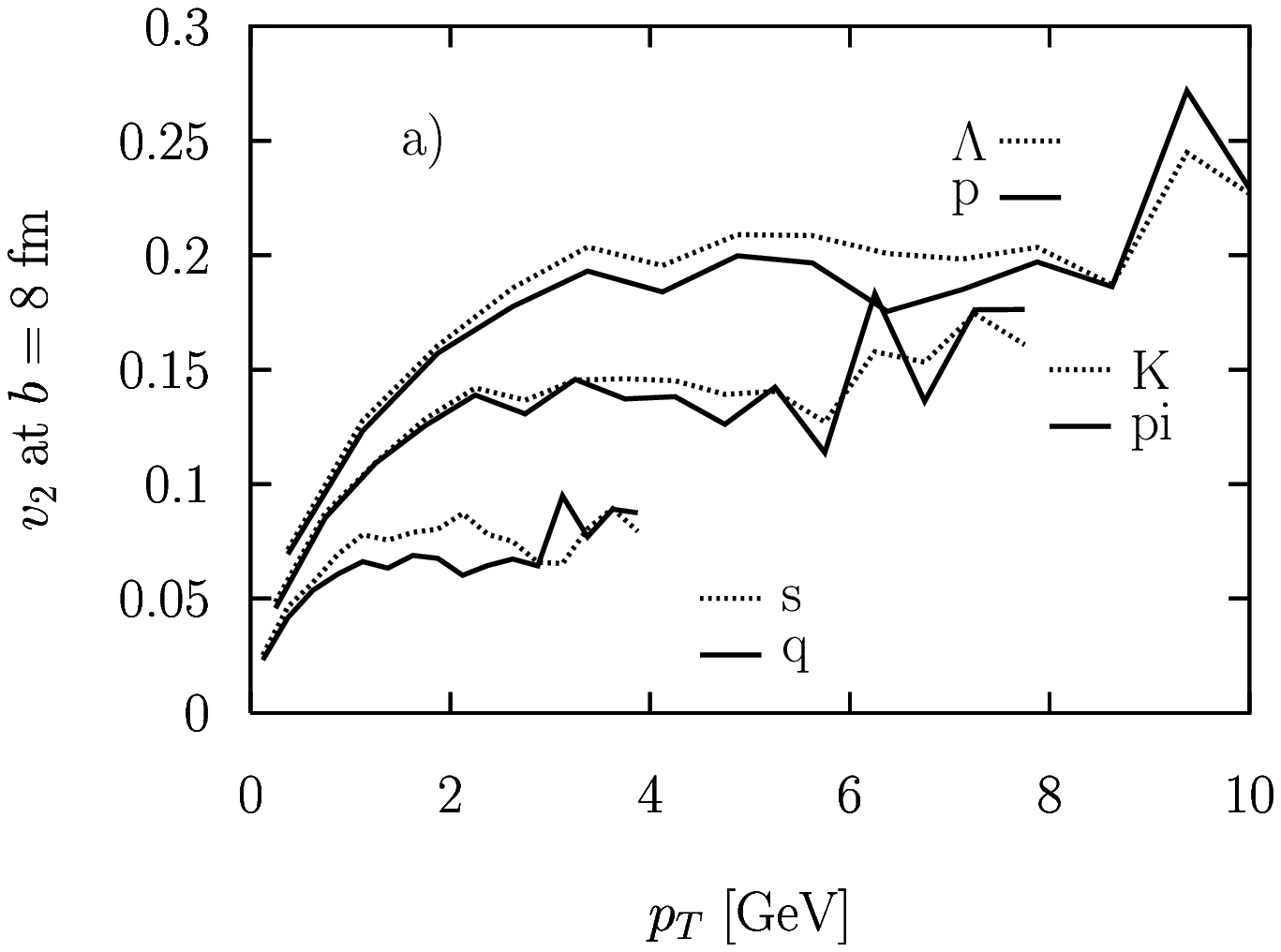,height=2.1in,width=2.7in,clip=5,angle=0} 
\epsfig{file=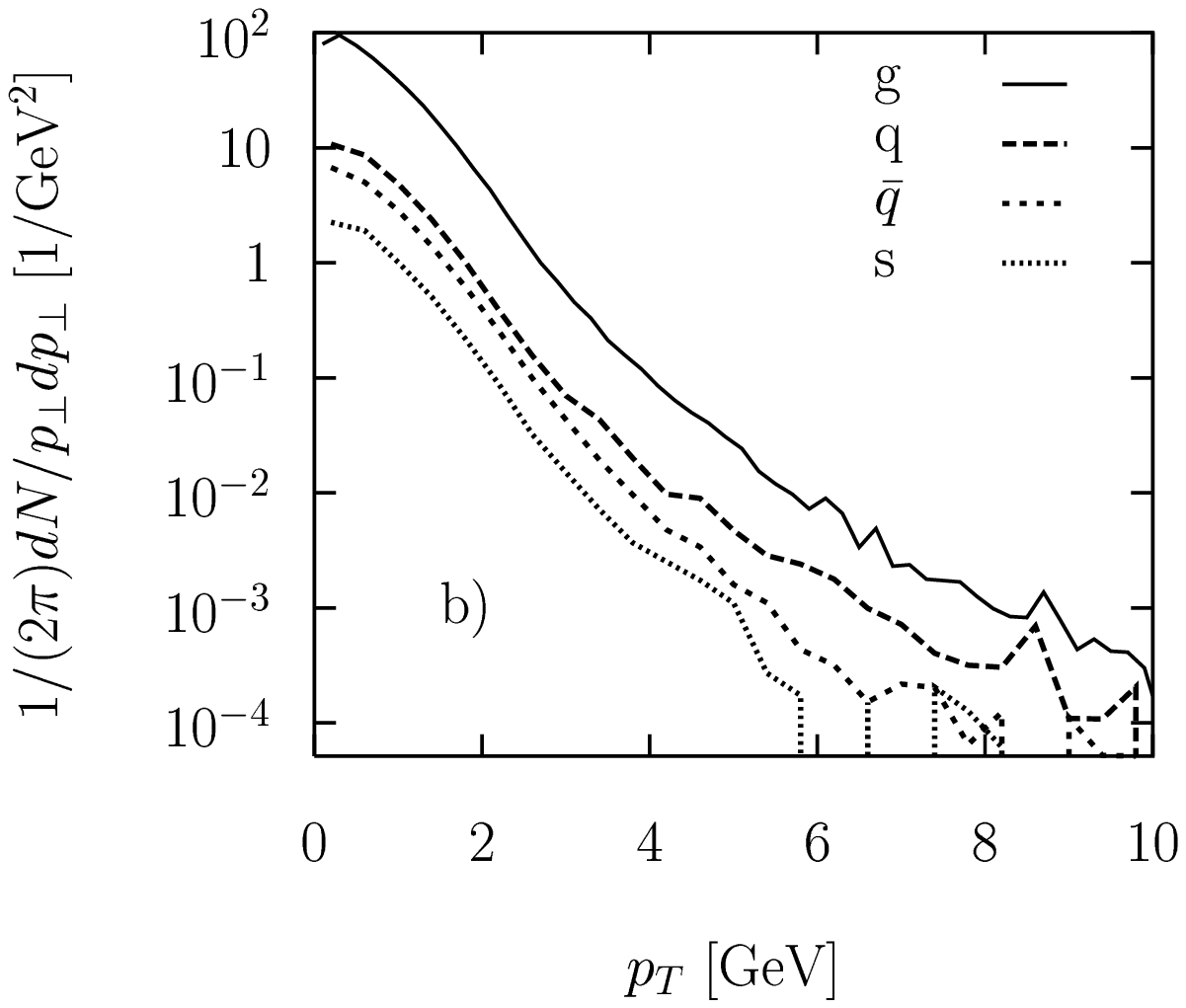,height=2.1in,width=2.4in,clip=5,angle=0}
\vspace*{-0.2cm} 
\caption{\label{fig3}
a) Preliminary elliptic flow ordering results from the MPC parton 
transport model\cite{MPC}.
b) Initial parton spectra for the MPC simulations.
}
\end{figure}

The above conclusions are also supported by
detailed numerical calculations using the MPC parton transport model\cite{MPC}.
Fig.~\ref{fig3}a shows the 
elliptic flow 
results for $Au+Au$ with $b=8$ fm at $\sqrt{s}=200A$ GeV at RHIC, 
computed from the azimuthal distribution of hadrons obtained
via Eq.~(\ref{coal_eq}).
The initial parton spectra, shown in Fig.~\ref{fig3}b, were computed from
leading-order perturbative QCD 
(GRV98LO, BKK95, $K=2.5$, $Q^2=p_\perp^2$) with a cutoff $p_0 = 2$ GeV,
below which the spectra were continued smoothly down to $p_\perp = 0$
to yield a $p_\perp$-integrated parton density $dN/d\eta = 2000$ at $\eta=0$.
The parton cross sections 
were $\sigma_{gg\to gg} = 3$ mb =  
$(9/4)\sigma_{qg\to qg} = (9/4)^2 \sigma_{qq\to qq}$.
The preliminary calculation shows rather small difference
between the elliptic flow of strange and light quarks.
Surprisingly, $v_2^s$ is slightly larger than $v_2^q$, however,
this may be because
flavor changing channels $gg\leftrightarrow qq$ and 
$q_i q_i\to q_j q_j$ were not included.

\section{Solution to the opacity puzzle}
While hadronization via $1 parton\to 1\pi$ or independent fragmentation
approximately preserves elliptic flow at high $2 < p_\perp < 6$ GeV 
\cite{v2},
parton coalescence enhances $v_2$ two times for mesons and three times for 
baryons.
Hence, the same hadron elliptic flow can be reached
from $2-3$ times smaller parton $v_2$,
i.e., with smaller parton densities and/or cross sections.

To determine the reduction of the parton opacity quantitatively,
one can utilize the results of Ref.~\cite{v2} that
computed gluon elliptic flow as a function of 
the transport opacity, 
$\chi\equiv \int dz\, \sigma_{tr} \rho(z) 
\approx \sigma dN/d\eta\,/\, (940$ mb$)$, from 
elastic parton transport theory for a minijet scenario of Au+Au at 
$\sqrt{s}= 130A$ GeV at RHIC.
Those results can be conveniently parameterized as 
$v_2(p_\perp,\chi) = v_2^{max}(\chi) \tanh[p_\perp / p_0(\chi)]$,
where $v_2^{max}$ is the saturation value of elliptic flow,
while $p_0$ is the $p_\perp$ scale above which saturation sets in.
For the estimates here 
one may assume that all gluons convert, e.g., via $gg\to qq$, 
to quarks of similar $p_\perp$ and hence $v_2^q(p_\perp) = v_2^g(p_\perp)$.

Because 
the increase of elliptic flow with opacity is weaker than 
linear\cite{v2,coalv2}, 
$v_2^{max}\sim \chi^{0.61}$,
the charged particle $v_2$ data from RHIC can be reproduced
in the coalescence scenario with $3-6$ times smaller parton opacities 
$\sigma dN/d\eta(b=0) \sim 7000-15000$ mb than those found in Ref.~\cite{v2}.
The lower(upper) value applies if high-$p_\perp$ hadrons are mostly
 baryons(mesons). 
Preliminary PHENIX data \cite{PHENIXhighpt} show
$\pi_0 / h^{\pm} \approx 0.5$ between $2 < p_\perp < 9 $ GeV,
therefore
one may expect $mesons/baryons\approx 1$,
in which case $\sigma dN/d\eta(b=0) \approx 10\,000$~mb.

In Ref.~\cite{v2} only collective flow was considered and
the parton opacity at RHIC was extracted using elliptic flow data from the
reaction plane analysis.
Taking into account non-flow effects that contributed
up to 15-20\% \cite{cumulantv2}
to the first elliptic flow measurements,
parton opacities should be further reduced by $25\%$
to $\sigma dN/d\eta(b=0) \sim 5000-10000$~mb.
For a typical elastic $gg\to gg$ cross section of $3$ mb,
this corresponds to an initial parton density $dN/d\eta(b=0) \sim 1500-3000$,
only $1.5-3$ times above the EKRT perturbative estimate\cite{EKRT}.

The remaining much smaller discrepancy is within
theoretical uncertainties.
For example, perturbative cross section and parton density
estimates may be too low.
If most hadrons form via coalescence,
the observed hadron multiplicity $dN_h/d\eta \approx 1000$ would imply
higher initial parton densities $dN/d\eta \sim 2000-3000$.
Constituent quark cross sections, $\sigma_{qq} \approx 4-5$ mb,
also point above the $\approx 3$ mb perturbative estimate.
One effect that estimate ignores is the enhancement of parton cross sections 
$\sigma \propto \alpha_s^2/\mu^2$
due to the decrease 
of the self-consistent Debye screening mass 
$\mu \sim gT_{eff}(\tau)$ during the expansion.
Finally, the contribution of inelastic processes,
such as $gg\leftrightarrow ggg$, to the opacity has also 
been neglected so far.
A preliminary study shows\cite{inelv2} that
this contribution can be similar to that of elastic processes.

\ack
Valuable comments by U.~Heinz and M.~Gyulassy 
and computer resources by the PDSF/LBNL are gratefully acknowledged. 
This work was supported 
by DOE grants DE-FG02-01ER41190 and DE-FG02-92ER40713.

\end{document}